\documentclass{PoS}
\usepackage{pstricks}

\title{Optimisation of Quantum Evolution Algorithms}

\ShortTitle{Optimisation of Quantum Evolution Algorithms}

\author{\speaker{Apoorva Patel}\\
        CHEP and SERC, Indian Institute of Science, Bangalore 560012, India\\
        E-mail: \email{adpatel@cts.iisc.ernet.in}}

\abstract{
Given a quantum Hamiltonian and its evolution time, the corresponding
unitary evolution operator can be constructed in many different ways,
corresponding to different trajectories between the desired end-points.
A choice among these trajectories can then be made to obtain the best
computational complexity and control over errors. As an explicit example,
Grover's quantum search algorithm is described as a Hamiltonian evolution
problem. It is shown that the computational complexity has a power-law
dependence on error when a straightforward Lie-Trotter discretisation
formula is used, and it becomes logarithmic in error when reflection
operators are used. The exponential change in error control is striking,
and can be used to improve many importance sampling methods. The key
concept is to make the evolution steps as large as possible while obeying
the constraints of the problem. In particular, we can understand why
overrelaxation algorithms are superior to small step size algorithms.
}

\FullConference{The 32nd International Symposium on Lattice Field Theory\\
                 23-28 June 2014\\
                 Columbia University, New York, NY}

\begin{document}

Classical computer simulations of quantum systems are not
efficient---well-known examples range from the Hubbard model to lattice
QCD---and Feynman argued that quantum simulations would do far better
\cite{feynman}. The essence of the argument is that quantum simulations
sum multiple evolutionary paths (in superposition) contributing to a
quantum process at one go, while classical simulations evaluate these
paths one by one. Formalisation of this advantage for simulating physical
Hamiltonians, in terms of computational complexity, has improved over the
years step by step \cite{lloyd,aharonov,BACS,BCCKS,inprog}. Here,
treating Grover's quantum search algorithm as a Hamiltonian evolution
problem, we expose the physical reasons behind the improvement in
computational complexity.

Computational complexity of a problem is a measure of the resources
needed to solve it. Conventionally, the computational complexity of a
decision problem is specified in terms of the size of its input, noting
that the size of its output is only one bit. Problems with different output
requirements are reduced to a sequence of decision problems, with gradually
narrowing bounds on the output adding one bit of precision for every
decision made. In such a scenario, the number of decision problems solved
equals the number of output bits, and it is appropriate to specify the
complexity of the original problem in terms of the size of its input as
well as its output. Generalising the conventional classification, the
computational algorithm then can be labeled efficient if the required
resources are polynomial in terms of the size of both its input and its
output. 

Popular importance sampling methods are not efficient according to our
criterion, because the number of iterations needed in the computational
effort has a negative power-law dependence on the precision $\epsilon$
(i.e. $N_{\rm iter}\propto\epsilon^{-2}$ as per the central limit theorem).
On the other hand, finding zeroes of a function by bisection is efficient
(i.e. $N_{\rm iter}\propto\log\epsilon$), and finding them by Newton's
method is super-efficient (i.e. $N_{\rm iter}\propto\log\log\epsilon$).

\section{Quantum Hamiltonian Simulation}

The Hamiltonian simulation problem is to evolve an initial quantum state
$|\psi(0)\rangle$ to a final quantum state $|\psi(T)\rangle$, in presence
of interactions specified by a Hamiltonian $H(t)$:
\begin{equation}
|\psi(T)\rangle = U(T) |\psi(0)\rangle ~,~~
U(T) =  P \Big[ \exp\big(-i\int_0^T H(t) dt\big) \Big] ~.
\end{equation}
Alternatively, the problem can be defined as determination of the
evolution operator $U(T)$, without any mention of the initial and the
final states. The norm of the difference between the simulated and the
exact evolution operators specifies the simulation accuracy, say
$||\widetilde{U}(T)-U(T)|| < \epsilon$.

We restrict ourselves here to Hamiltonians acting in finite $N$-dimensional
Hilbert spaces. A general $H(t)$ then be a dense $N \times N$ matrix, and
there is no efficient way to simulate it. So we furthermore assume that
$H(t)$ the following features commonly present in physical problems:\\
(1) The Hilbert space is a tensor product of many components, e.g.
$N=2^m$ for a system of qubits.\\
(2) The components have only local interactions irrespective of the
size of the system, e.g. only nearest neighbour couplings. That makes
$H(t)$ sparse, with $O(N)$ non-zero elements.\\
(3) $H(t)$ is specified in terms of a finite number of functions, while
the arguments of the functions can depend on the components, e.g. the
interactions are translationally invariant. That allows $H(t)$ to have
a compact description, and consequently the resources needed to just
write down $H(t)$ do not influence the simulation complexity.

Such Hamiltonians can be mapped to graphs with bounded degree $d$,
with vertices $\leftrightarrow$ components and edges $\leftrightarrow$
interactions. Their simulations can be easily parallelised---on classical
computers, they allow SIMD simulations with domain decomposition. With
these criteria, efficient Hamiltonian simulation algorithms are those
that use resources polynomial in $\log(N)$, $d$ and $\log(\epsilon)$.

\subsection{Hamiltonian Decomposition}

Efficient simulation strategy for Hamiltonian evolution has two major
ingredients. The first ingredient is to decompose the sparse Hamiltonian
as a sum of non-commuting but block-diagonal Hermitian operators, i.e.
$H =\sum_{i=1}^l H_i$. Then each $H_i$ can be easily and exactly
exponentiated for any time evolution $\tau$, with $\exp(-iH_i\tau)$
retaining the same block-diagonal structure. Reducing the block size
all the way to $2\times2$, the blocks become linear combinations of
projection operators. Projection operators with only two distinct
eigenvalues can be interpreted as binary query oracles.

In general, $H_i$ can be identified by an edge-colouring algorithm for
graphs \cite{aharonov}, with distinct colours for overlapping edges.
At most $d+1$ colours are needed to efficiently colour any sparse graph.
Identification of $H_i$ also provides a compressed labeling scheme that
can be used to address individual blocks. The number of blocks is
$O(m)=O(\log N)$, and they can be evolved simultaneously, in parallel
(classically) or in superposition (quantum mechanically).

For example, even and odd edges of a linear chain provide a block-diagonal
decomposition of the one-dimensional Laplacian operator, $H=H_o+H_e$.
Its projection operator structure follows from $H_o^2=2H_o$ and $H_e^2=2H_e$.
The last bit of the position label identifies $H_o$ and $H_e$. Eigenvalues
of $H$ are $4\sin^2(k/2)$ in terms of the lattice momentum $k$, while those
of $H_o$ and $H_e$ are just $0$ and $2$.

\subsection{Evolution Optimisation}

Given that individual $H_i$ can be exponentiated exactly and efficiently,
their sum $H$ can be approximately but efficiently exponentiated using
the discrete Lie-Trotter formula:
\begin{equation}
\label{LieTrotter}
\exp\big(-i H T\big) = \exp\Big(-i\sum_i H_i T\Big)
  \approx \Big(\prod_i \exp(-iH_i \Delta t)\Big)^n ~,~~ n = T/\Delta t ~.
\end{equation}
This approximation retains unitarity of the evolution, but may not
preserve other properties such as the energy. The accuracy of the
approximation is usually improved by decreasing $\Delta t$. This
method has been used in classical parallel computer simulations of
quantum evolution problems \cite{deraedt,richardson}.

In contrast, the second ingredient of efficient Hamiltonian simulation is
to use as large $\Delta t$ as possible. When the exponent is proportional
to a projection operator, the largest $\Delta t$ is the one that makes
the exponential a reflection operator. Such an extreme strategy not only
keeps the evolution accurate but also improves the algorithmic complexity
from a power-law dependence on $\epsilon$ to a logarithmic one. This is
not obvious, and we demonstrate it next for the quantum search problem.

\section{Quantum Search as Hamiltonian Evolution}

The quantum search algorithm works in an $N$-dimensional Hilbert space,
whose basis vectors $\{|i\rangle\}$ are identified with the individual items.
It takes the initial state $|s\rangle$ whose amplitudes are uniformly
distributed over all the items, to the target state $|t\rangle$ where
all but one amplitudes vanish.
\begin{equation}
|\psi(0)\rangle = |s\rangle ~,~~ |\psi(T)\rangle = |t\rangle ~,~~
|\langle i|s \rangle| = 1/\sqrt{N} ~,~~ \langle i|t \rangle = \delta_{it} ~.
\end{equation}

The simplest evolution schemes taking $|s\rangle$ to $|t\rangle$ are
governed by time-independent Hamiltonians that depend only on $|s\rangle$
and $|t\rangle$. The unitary evolution is then a rotation at a fixed rate
in the two-dimensional subspace, formed by $|s\rangle$ and $|t\rangle$,
of the whole Hilbert space. In this subspace, let
\begin{equation}
|t\rangle = \pmatrix{1 \cr 0} ~,~~ |t_\perp\rangle = \pmatrix{0 \cr 1} ~,~~
|s\rangle = \pmatrix{1/\sqrt{N} \cr \sqrt{(N-1)/N}} ~.
\end{equation}
There are many evolution routes with $U(T)|s\rangle=|t\rangle$, and we
consider two particular cases in turn.

\subsection{Farhi-Gutmann's and Grover's Algorithms}

Grover based his algorithm on a physical intuition for the Hamiltonian
\cite{trotter}, where the potential energy term $|t\rangle\langle t|$
attracts the wavefunction towards the target state and the kinetic energy
term $|s\rangle\langle s|$ diffuses the wavefunction over the whole Hilbert
space. Both the terms are projection operators, and the time-independent
Hamiltonian is
\begin{equation}
\label{HamFG}
H_C = |s\rangle\langle s| + |t\rangle\langle t|
    = I + {\sqrt{N-1}\over N} \sigma_1 + {1\over N} \sigma_3 ~.
\end{equation}
The corresponding evolution operator is (without the global phase) 
\begin{equation}
U_C(t) = \exp\big(-i\hat{n}\cdot\vec{\sigma}~t/\sqrt{N}\big) ~,~~
\hat{n} = \big(\sqrt{(N-1)/N}, 0, 1/\sqrt{N}\big)^T ~,
\end{equation}
which is a rotation by angle $2t/\sqrt{N}$ around the direction defined
by $\hat{n}$ on the Bloch sphere.

The (unnormalised) eigenvectors of $H_C$ are $|s\rangle\pm|t\rangle$,
They correspond to the directions $\pm\hat{n}$, and bisect the initial
and the target states. Thus a rotation by angle $\pi$ around $\hat{n}$
takes $|s\rangle\langle s|$ to $|t\rangle\langle t|$ on the Bloch sphere,
and the time required for the Hamiltonian search is $T=(\pi/2)\sqrt{N}$
\cite{farhi}.

Grover made an enlightened jump from this scenario, motivated by the
Lie-Trotter formula. He exponentiated the projection operators in $H_C$
to reflection operators; $R=\exp(\pm i\pi P)=1-2P$ for any projection
operator $P$. His optimal algorithm iterates the discrete evolution
operator \cite{grover},
\begin{equation}
\label{operG}
U_G = -(1-2|s\rangle\langle s|)(1-2|t\rangle\langle t|)
    = (1-{2\over N}) I + 2i{\sqrt{N-1}\over N} \sigma_2 ~.
\end{equation}
With $U_G = \exp(-iH_G\tau)$, it corresponds to the Hamiltonian
and the evolution step:
\begin{equation}
H_G = {i\over\sqrt{N}} \big(|t\rangle\langle s| - |s\rangle\langle t|\big)
    = i\big[|t\rangle\langle t|,|s\rangle\langle s|\big]
    = -{\sqrt{N-1}\over N} \sigma_2 ~,~~
\tau = {2N\over\sqrt{N-1}} \sin^{-1}\Big( {1\over\sqrt{N}} \Big) .
\end{equation}
It is an important non-trivial fact that $H_G$ is the commutator
of the two projection operators in $H_C$.

On the Bloch sphere, each $U_G$ step is a rotation by angle
$2\tau\sqrt{N-1}/N = 4\sin^{-1}(1/\sqrt{N})$ around the direction
$\hat{n}_G=(0,1,0)^T$, taking the geodesic route from the initial
to the final state. That makes the number of steps required for this
discrete Hamiltonian search,
\begin{equation}
Q_T = {\cos^{-1}(1/\sqrt{N}) \over 2\sin^{-1}(1/\sqrt{N})}
    \approx {\pi\over4}\sqrt{N} ~.
\end{equation}

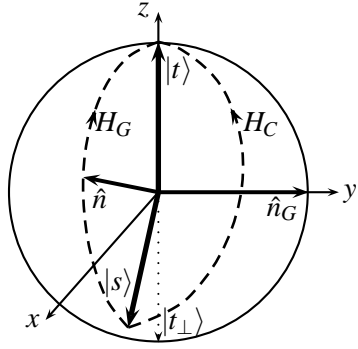
\begin{figure}[t]
{
\begin{center}
\begin{picture}(250,120)
% Unit length is 1cm for pstricks
\psline{->}(4,2)(6.4,2)
\put(184,55){$y$}
\psline{->}(4,2)(4,4.4)
\put(106,124){$z$}
\psline{->}(4,2)(2.5,0.3)
\put(64,6){$x$}

\psline[linewidth=2pt]{->}(4,2)(3.6,0.2)
\put(93,20){$|s\rangle$}
\psline[linewidth=2pt]{->}(4,2)(4,4)
\put(116,100){$|t\rangle$}
\psline[linestyle=dotted]{->}(4,2)(4,0)
\put(115,5){$|t_\perp\rangle$}

\psline[linewidth=1.5pt]{->}(4,2)(6,2)
\put(155,48){$\hat{n}_G$}
\psline[linewidth=1.5pt]{->}(4,2)(3,2.2)
\put(89,51){$\hat{n}$}

\pscircle(4,2){2}
%\psellipse[linestyle=dotted](4,2)(1,2)
%\psarc[linestyle=dotted](3.03,2.18){2.06}{286}{62}
\pscurve[linestyle=dashed,linewidth=1pt](3.6,0.2)(3.3,0.6)(3,2)(3.2,3.2)(3.6,3.8)(4,4)
\pscurve[linestyle=dashed,linewidth=1pt](3.6,0.2)(4.4,0.6)(5.1,2)(4.9,3.3)(4.3,3.9)(4,4)

\put(90,80){$H_G$}
\psline[linewidth=1.5pt]{->}(3.15,3.06)(3.16,3.1)
\put(146,80){$H_C$}
\psline[linewidth=1.5pt]{->}(4.98,3.08)(4.96,3.12)
\end{picture}
\end{center}
}
\vspace{-4mm}
\caption{Evolution trajectories on the Bloch sphere for the quantum
search problem, going from $|s\rangle$ to $|t\rangle$. The Hamiltonians
$H_C$ and $H_G$ generate rotations around the directions $\hat{n}$ and
$\hat{n}_G$ respectively.}
\label{evoltraj}
\end{figure}

Note that $\hat{n}$ and $\hat{n}_G$ are orthogonal, so the evolution
trajectories produced by rotations around them are completely different,
as illustrated in Fig.\ref{evoltraj}. It is only after a specific
evolution time, corresponding to the solution of the quantum search
problem, that the two trajectories meet.

To compare the rates of these two Hamiltonian evolutions, we observe that
$H_C$ can be simulated by alternating small evolution steps governed by
$|s\rangle\langle s|$ and $|t\rangle\langle t|$, according to the Lie-Trotter
formula. Then each evolution step governed by $|t\rangle\langle t|$ needs
two binary queries \cite{nielsen}. On the other hand, $U_G$ can be
simulated using only one binary query per evolution step.

\subsection{Equivalent Evolutions}

Two Hamiltonian evolutions are truly equivalent, when their corresponding
unitary evolution operators are the same (upto a global phase). The
intersection of the two evolution trajectories is then independent of
the specific initial and final states. For the quantum search problem,
we find
\begin{equation}
\label{equivevolint}
U_C(T) = i (1-2|t\rangle\langle t|) ~ (U_G)^{Q_T} ~.
\end{equation}
For a general evolution time $0<t<T$, we have the relation
(similar to Euler angle decomposition),
\begin{equation}
\label{equivevolfrac}
U_C(t) = \exp\big(i\beta\sigma_3\big) ~(U_G)^{Q_t}~
         \exp\Big(i\big({\pi\over2}+\beta\big)\sigma_3\Big) ~,
\end{equation}
i.e. $U_C(t)$ can be generated as $Q_t$ iterations of the Grover
operator $U_G$, preceded and followed by phase rotations. Here
$\sigma_3=2|t\rangle\langle t|-1$ is a known reflection, and
\begin{equation}
\label{evolfracparam}
Q_t = {\sin^{-1}\Big(\sqrt{N-1\over N} \sin(t/\sqrt{N})\Big)
      \over 2\sin^{-1}(1/\sqrt{N})} \approx {t\over2} ~,~~
\beta = -{\pi\over4}
        -{1\over2}\tan^{-1}\Big({1\over\sqrt{N}}\tan(t/\sqrt{N})\Big) ~.
\end{equation}

It is truly remarkable that $H_G$ can be used to obtain the same evolution
as $H_C$, even though the two Hamiltonians are entirely different in terms
of their eigenvectors and eigenvalues!

\subsection{Discretised Hamiltonian Evolution Complexity}

Digitisation of continuous variables is necessary for fault-tolerant
computation with control over bounded errors. But it also introduces
discretisation errors that must be kept within specified bounds.
The algorithmic error of the Lie-Trotter formula depends on $\Delta t$,
which has to be chosen so as to satisfy the total error bound $\epsilon$
on $U(t)$. For the simplest discretisation scheme, 
\begin{equation}
\exp\Big(-i\sum_{i=1}^l H_i \Delta t\Big)
  = \exp\big(-iH_1 \Delta t\big) \ldots \exp\big(-iH_l \Delta t\big)
  \times \exp\big(-iE^{(2)} (\Delta t)^2 \big) ~,
\end{equation}
\begin{equation}
E^{(2)} = {i\over2} \sum_{i<j} [H_i,H_j] + O(\Delta t) ~.
\end{equation}
For unitary operators $X$ and $Y$, Cauchy-Schwarz and triangle inequalities
give,
\begin{equation}
||X^n - Y^n|| = ||(X-Y)(X^{n-1}+\ldots+Y^{n-1})|| \le n||X-Y|| ~.
\end{equation}
So for the total evolution to remain within the error bound $\epsilon$,
we need
\begin{equation}
n ||\exp(-iE^{(2)} (\Delta t)^2) - I|| \approx n ||E^{(2)}|| (\Delta t)^2
  = t ||E^{(2)}|| (\Delta t) < \epsilon ~.
\end{equation}
With exact exponentiation of the individual terms $H_i$, the computational
cost to simulate a single time step $\Delta t$, $\cal C$, does not depend
on $\Delta t$. The complexity of the Hamiltonian evolution is then
\begin{equation}
O(n {\cal C}) = O\Big(t^2
  \Big({||E^{(2)}||\over\epsilon}\Big) {\cal C}\Big) ~.
\end{equation}
With superlinear scaling in $t$ and power-law scaling in $\epsilon$,
this small $\Delta t$ scheme is not efficient.

Grover's optimal algorithm uses a discretisation formula where
$\exp(-iH_i \Delta t_G)$ are reflection operators. The corresponding
time step is large, i.e. $\Delta t_G = \pi$ for Eq.(\ref{LieTrotter})
applied to Eq.(\ref{HamFG}). The large time step introduces an error
because one may jump across the target state during evolution
instead of reaching it exactly. $Q_t$ is not an integer as defined
in Eq.(\ref{evolfracparam}), and needs to be replaced by its nearest
integer approximation $\lfloor Q_t+{1\over2}\rfloor$ in practice.
Since each time step provides a rotation by angle
$\alpha=2\sin^{-1}(1/\sqrt{N})$, and one may miss the target state by
at most half a rotation step, the error probability of Grover's algorithm is
bounded by $\sin^2(\alpha/2)=1/N$, independent of the number of time steps.
Since the preceding and following phase rotations in Eq.(\ref{equivevolfrac})
are unitary operations, this error bound applies to $U_C(t)$ as well.
Thereafter, multiple runs of the algorithm and selection of the result by
majority rule can rapidly reduce the error probability. With $R$ runs,
the error probability becomes less than $2^{R-1}/N^{\lceil R/2 \rceil}$,
which can be made smaller than any prescribed error bound $\epsilon$.
(In a drastic contrast, averaging the results of multiple runs would make
the error probability smaller than $1/(N\sqrt{R})$ only.) The computational
complexity of the evolution is thus
\begin{equation}
O(Q_t R {\cal C}_G) = O\Big({t\over 2}
                       \Big(-{2\log\epsilon\over\log N}\Big) {\cal C}_G\Big)
                    = O\Big(-t{\log\epsilon\over\log N} {\cal C}_G\Big) ~.
\end{equation}
With linear scaling in time and logarithmic scaling in $\epsilon$, 
this algorithm is efficient.

To complete the analysis, we note that a digital computer with a finite
register size also produces truncation errors. With $b$-bit registers,
the available precision is $\delta=2^{-b}$. With all functions approximated
by accurate polynomials, and Euler angle decomposition reducing rotations
about arbitrary axes to rotations about fixed axes, an individual $H_i$
can be exponentiated to $b$-bit precision with $O(mb^3)$ effort. The number
of exponentiations of $H_i$ needed for the Lie-Trotter formula is $nl$,
which reduces to $2Q_t$ for the Grover version. So with the choice
$nl\delta=O(\epsilon)$, i.e. $b=\Theta(\log(n/\epsilon))$, the truncation
error becomes negligible compared to the discretisation error. The cost of
a single evolution step then scales as ${\cal C}=O(m(\log(t/\epsilon))^3)$,
which is efficient.

\section{Extensions and Outlook}

It is straightforward to extend the preceding results to other Hamiltonians
consisting of only two projection operators, e.g. the staggered Dirac
operator for free fermions in any number of dimensions \cite{fermion_eo}.
For Hamiltonians that are linear combinations of more than two projection
operators, e.g. three projection operators for the discretised Laplacian
on the graphene lattice, successive $H_i$ can be added to the algorithm
one by one in an inductive procedure. The resultant large $\Delta t$
evolution is not exact, but it still has $\Theta(1)$ success probability
for a suitable choice of $\Delta t$. That keeps the overall scaling of the
evolution efficient, $O\big(lt||H||\log(lt||H||/\epsilon){\cal C}\big)$
\cite{BCCKS,inprog}.

Our construction of the efficient Hamiltonian evolution algorithm relies
on: (1) simplification of the Baker-Campbell-Hausdorff expansion for
products of exponentials of projection operators, and (2) conversion of
the results to a digital form allowing selection of the best one by
majority rule. These algebraic properties are not specific to quantum
computers; they can be incorporated in and readily benefit traditional
classical simulations of quantum systems. In particular:\\
(a) It is known that overrelaxation algorithms \cite{overrelax},
based on evolution steps that are reflections consistent with conservation
laws, provide a much more efficient sampling of the configuration space
(measured in terms of the autocorrelation time) than the small step size
Metropolis algorithm. The analysis presented here provides an understanding
of that observation.\\
(b) Many physical problems with periodic patterns are solved using the
fast Fourier transform. The block-diagonal decomposition provides a
competitive real space method for solving them.\\
(c) Projection operator decomposition of the Hamiltonian can be easily
found for quantum Monte Carlo problems, and for molecular dynamics
simulations using the Lie-Trotter formula in Euclidean time. The remaining
task is to find a useful large step size with high acceptance probability.\\
(d) Evaluations of functions other than exponentials may also simplify with
the block-diagonal projection operator decomposition. A case of particular
interest is the evaluation of the fermion determinant appearing in many
physical problems.

Work on such applications is in progress.

\end{document}